\documentstyle[prb,aps,twocolumn]{revtex}

\begin{document}

\draft
\flushbottom
\twocolumn[\hsize\textwidth\columnwidth\hsize\csname @twocolumnfalse\endcsname

\title{\bf Selfconsistent order-$N$ density-functional calculations
for very large systems}

\author{Pablo Ordej\'on,$^{1,2}$ Emilio Artacho,$^3$ and Jos\'e M. Soler$^3$}
\address{
$^1$Department of Physics and Materials Research Laboratory,
University of Illinois, Urbana, Illinois 61801, USA\\
$^2$Departamento de F{\'{\i}}sica,
Universidad de Oviedo, C/ Calvo Sotelo s/n, 33007 Oviedo, Spain\\
$^3$Instituto de Ciencia de Materiales Nicol\'as Cabrera and
Departamento de F{\'{\i}}sica de Materia Condensada, C-III \\
Universidad Aut\'onoma de Madrid, 28049 Madrid, Spain.}

\date{\today}
\maketitle

\begin{abstract}
We present a method to perform fully selfconsistent 
density-functional calculations, which scales linearly 
with the system size and which is well suited for very large systems. 
It uses strictly localized pseudoatomic orbitals as basis functions.
The sparse Hamiltonian and overlap matrices are calculated with an 
$O(N)$ effort.
The long range selfconsistent potential and its matrix elements
are computed in a real-space grid. The other matrix elements
are directly calculated and tabulated as a function of the
interatomic distances. 
The computation of the total energy
and atomic forces is also done in $O(N)$ operations using truncated, 
Wannier-like localized functions to describe the occupied states, 
and a band-energy functional which is iteratively minimized
with no orthogonality constraints. We illustrate the method 
with several examples, including carbon and
silicon supercells with up to 1000 Si
atoms and  supercells of $\beta$-C$_3$N$_4$.
We apply the method to solve the existing controversy
about the faceting of large icosahedral fullerenes by performing
dynamical simulations on C$_{60}$, C$_{240}$, and C$_{540}$. 
\end{abstract}

%\pacs{PACS numbers: 71.10.+x, 71.20.Ad, 61.46.+w}
\pacs{ }
]

% INTRODUCTION

A large effort has been devoted in the last few years to
develop approximate methods to solve the electronic structure of large
systems with a computational cost proportional to its 
size.\cite{ordejon}
Several approaches are now sufficiently accurate and robust 
to obtain reliable results for systems with 
thousands of atoms. So far, however, most of these schemes have 
been useful only with simple Hamiltonians, like
empirical tight-binding models, which provide an ideal
setting for order-$N$ calculations. 
%since the computation 
%of the Hamiltonian matrix elements is very simple and fast,
%and the minimal basis sets yield matrices of manageable  size.
First-principles order-$N$ calculations have been performed
mainly in the non-selfconsistent Harris functional 
version\cite{harris} of the local density approximation (LDA)
for electronic exchange and correlation (XC) using
minimal bases.\cite{ordejon,yang}
Linear scaling algorithms in fully selfconsistent LDA have also
been tried,\cite{hierse} but the results are far from the linear scaling 
regime, due to the relatively small number of manageable atoms in those
simulations. Hernandez {\it et al.}\cite{hernandez} have successfully 
produced LDA  results in large silicon systems using a real-space grid 
method. The computational requirements that this kind of approach 
demands are, however, extremely large, and calculations must be 
performed in massive computational platforms. 

We have developed a selfconsistent density-functional 
formulation with linear scaling, capable of producing results 
for very large systems, whose computational demands
are not overwhelmingly large, so that systems with many hundreds 
of atoms can be treated in modest computational platforms like 
workstations, and much larger systems can be treated in massive platforms.
The method is based on the linear combination
of atomic orbitals (LCAO) approximation as basis of expansion of
the electronic states. Non-orthogonal LCAO bases are very efficient, 
reducing the number of variables dramatically, 
compared to plane-wave (PW) or real-space-grid
approaches, so that larger systems can be studied. Also, LCAO can provide
up to extremely accurate bases, staying always in the range of a few 
valence orbitals per atom.\cite{poirier} As a first step, in this work we
use minimal basis sets of one $s$ and three $p$ orbitals
per atom, the extension to larger bases being perfectly possible
within the present formulation. 
The choice of a basis obviously implies a possible error associated to its
incompleteness. In the same way as for the error due to the linear 
scaling algorithm, the error in the basis can be reduced at the expense 
of an increase in computational effort. Its magnitude should be carefully 
checked, but also compared with other sources of error to ensure that an 
increase of the basis is really worthwhile.

% METHOD
 
Our method uses standard LDA techniques\cite{lda} for the valence electrons,
the core electrons being replaced by pseudopotentials.\cite{bachelet}
The basis orbitals used in this work are the 
$s$ and $p$ pseudoatomic orbitals defined by
Sankey and Niklewski,\cite{sankey}
in the context of non-selfconsistent Harris
functional methods. These are slightly excited pseudoatomic
orbitals $\phi_{\mu}({\bf r})$, obtained by solving the valence
electron problem in the isolated atom, with the same pseudopotential
and LDA approximations, and with the boundary condition
that the atomic orbitals are strictly localized,
vanishing outside a given radius $r_c$.
This radius cutoff is in principle orbital dependent, but  we 
do not make explicit this dependence in the equations only for 
simplicity in the notation.
The great advantage of these orbitals is that they give rise to
sparse overlap and Hamiltonian matrices (since matrix 
elements between distant orbitals exactly vanish exactly) 
and they display the 
same structure as in conventional tight-binding. The extent of the
interactions and the sparseness of the matrices depend on
the cutoff radius $r_c$ of each atom. These are not critical
as long as the maxima of the atomic wave functions are well within $r_c$. 
For an analysis of the quality of pseudoatomic
orbitals as a basis for solid state calculations we refer the
reader to Ref.~\onlinecite{bases}.

\begin{figure}
\parbox{8.3cm}{ \vspace{-0.8cm} \hspace{-0.8cm}
%\input{figc64}
% GNUPLOT: LaTeX picture
\setlength{\unitlength}{0.240900pt}
\ifx\plotpoint\undefined\newsavebox{\plotpoint}\fi
\sbox{\plotpoint}{\rule[-0.175pt]{0.350pt}{0.350pt}}%
\begin{picture}(1049,720)(0,0)
\tenrm
\sbox{\plotpoint}{\rule[-0.175pt]{0.350pt}{0.350pt}}%
\put(264,214){\rule[-0.175pt]{4.818pt}{0.350pt}}
\put(242,214){\makebox(0,0)[r]{0.0}}
\put(965,214){\rule[-0.175pt]{4.818pt}{0.350pt}}
\put(264,326){\rule[-0.175pt]{4.818pt}{0.350pt}}
\put(242,326){\makebox(0,0)[r]{0.1}}
\put(965,326){\rule[-0.175pt]{4.818pt}{0.350pt}}
\put(264,439){\rule[-0.175pt]{4.818pt}{0.350pt}}
\put(242,439){\makebox(0,0)[r]{0.2}}
\put(965,439){\rule[-0.175pt]{4.818pt}{0.350pt}}
\put(264,551){\rule[-0.175pt]{4.818pt}{0.350pt}}
\put(242,551){\makebox(0,0)[r]{0.3}}
\put(965,551){\rule[-0.175pt]{4.818pt}{0.350pt}}
\put(264,158){\rule[-0.175pt]{0.350pt}{4.818pt}}
\put(264,113){\makebox(0,0){0}}
\put(264,587){\rule[-0.175pt]{0.350pt}{4.818pt}}
\put(428,158){\rule[-0.175pt]{0.350pt}{4.818pt}}
\put(428,113){\makebox(0,0){25}}
\put(428,587){\rule[-0.175pt]{0.350pt}{4.818pt}}
\put(592,158){\rule[-0.175pt]{0.350pt}{4.818pt}}
\put(592,113){\makebox(0,0){50}}
\put(592,587){\rule[-0.175pt]{0.350pt}{4.818pt}}
\put(756,158){\rule[-0.175pt]{0.350pt}{4.818pt}}
\put(756,113){\makebox(0,0){75}}
\put(756,587){\rule[-0.175pt]{0.350pt}{4.818pt}}
\put(919,158){\rule[-0.175pt]{0.350pt}{4.818pt}}
\put(919,113){\makebox(0,0){100}}
\put(919,587){\rule[-0.175pt]{0.350pt}{4.818pt}}
\put(264,158){\rule[-0.175pt]{173.689pt}{0.350pt}}
\put(985,158){\rule[-0.175pt]{0.350pt}{108.164pt}}
\put(264,607){\rule[-0.175pt]{173.689pt}{0.350pt}}
\put(23,382){\makebox(0,0)[l]{\shortstack{\large $\Delta E$ (eV)}}}
\put(624,68){\makebox(0,0){\large $E_{cut}$ (Ry)}}
\put(264,158){\rule[-0.175pt]{0.350pt}{108.164pt}}
\sbox{\plotpoint}{\rule[-0.500pt]{1.000pt}{1.000pt}}%
\put(499,565){\usebox{\plotpoint}}
\put(499,562){\usebox{\plotpoint}}
\put(500,559){\usebox{\plotpoint}}
\put(501,556){\usebox{\plotpoint}}
\put(502,553){\usebox{\plotpoint}}
\put(503,550){\usebox{\plotpoint}}
\put(504,548){\usebox{\plotpoint}}
\put(505,545){\usebox{\plotpoint}}
\put(506,542){\usebox{\plotpoint}}
\put(507,539){\usebox{\plotpoint}}
\put(508,536){\usebox{\plotpoint}}
\put(509,534){\usebox{\plotpoint}}
\put(510,531){\usebox{\plotpoint}}
\put(511,528){\usebox{\plotpoint}}
\put(512,525){\usebox{\plotpoint}}
\put(513,522){\usebox{\plotpoint}}
\put(514,519){\usebox{\plotpoint}}
\put(515,517){\usebox{\plotpoint}}
\put(516,514){\usebox{\plotpoint}}
\put(517,511){\usebox{\plotpoint}}
\put(518,508){\usebox{\plotpoint}}
\put(519,505){\usebox{\plotpoint}}
\put(520,503){\usebox{\plotpoint}}
\put(521,500){\usebox{\plotpoint}}
\put(522,497){\usebox{\plotpoint}}
\put(523,494){\usebox{\plotpoint}}
\put(524,491){\usebox{\plotpoint}}
\put(525,489){\usebox{\plotpoint}}
\put(526,489){\usebox{\plotpoint}}
\put(527,488){\usebox{\plotpoint}}
\put(528,487){\usebox{\plotpoint}}
\put(530,486){\usebox{\plotpoint}}
\put(531,485){\usebox{\plotpoint}}
\put(532,484){\usebox{\plotpoint}}
\put(534,483){\usebox{\plotpoint}}
\put(535,482){\usebox{\plotpoint}}
\put(537,481){\usebox{\plotpoint}}
\put(538,480){\usebox{\plotpoint}}
\put(539,479){\usebox{\plotpoint}}
\put(541,478){\usebox{\plotpoint}}
\put(542,477){\usebox{\plotpoint}}
\put(543,476){\usebox{\plotpoint}}
\put(545,475){\usebox{\plotpoint}}
\put(546,474){\usebox{\plotpoint}}
\put(548,473){\usebox{\plotpoint}}
\put(549,472){\usebox{\plotpoint}}
\put(550,471){\usebox{\plotpoint}}
\put(552,470){\usebox{\plotpoint}}
\put(553,469){\usebox{\plotpoint}}
\put(555,468){\usebox{\plotpoint}}
\put(558,467){\usebox{\plotpoint}}
\put(562,466){\usebox{\plotpoint}}
\put(566,465){\usebox{\plotpoint}}
\put(570,464){\usebox{\plotpoint}}
\put(573,463){\usebox{\plotpoint}}
\put(577,462){\usebox{\plotpoint}}
\put(581,461){\usebox{\plotpoint}}
\put(585,460){\usebox{\plotpoint}}
\put(587,459){\usebox{\plotpoint}}
\put(589,458){\usebox{\plotpoint}}
\put(591,457){\usebox{\plotpoint}}
\put(593,456){\usebox{\plotpoint}}
\put(595,455){\usebox{\plotpoint}}
\put(597,454){\usebox{\plotpoint}}
\put(599,453){\usebox{\plotpoint}}
\put(602,452){\usebox{\plotpoint}}
\put(604,451){\usebox{\plotpoint}}
\put(606,450){\usebox{\plotpoint}}
\put(608,449){\usebox{\plotpoint}}
\put(610,448){\usebox{\plotpoint}}
\put(612,447){\usebox{\plotpoint}}
\put(614,446){\usebox{\plotpoint}}
\put(617,445){\usebox{\plotpoint}}
\put(618,444){\usebox{\plotpoint}}
\put(619,443){\usebox{\plotpoint}}
\put(620,442){\usebox{\plotpoint}}
\put(621,441){\usebox{\plotpoint}}
\put(622,440){\usebox{\plotpoint}}
\put(623,439){\usebox{\plotpoint}}
\put(624,438){\usebox{\plotpoint}}
\put(625,437){\usebox{\plotpoint}}
\put(626,436){\usebox{\plotpoint}}
\put(628,435){\usebox{\plotpoint}}
\put(629,434){\usebox{\plotpoint}}
\put(630,433){\usebox{\plotpoint}}
\put(631,432){\usebox{\plotpoint}}
\put(632,431){\usebox{\plotpoint}}
\put(633,430){\usebox{\plotpoint}}
\put(634,429){\usebox{\plotpoint}}
\put(635,428){\usebox{\plotpoint}}
\put(636,427){\usebox{\plotpoint}}
\put(637,426){\usebox{\plotpoint}}
\put(639,425){\usebox{\plotpoint}}
\put(640,424){\usebox{\plotpoint}}
\put(641,423){\usebox{\plotpoint}}
\put(642,422){\usebox{\plotpoint}}
\put(643,421){\usebox{\plotpoint}}
\put(644,420){\usebox{\plotpoint}}
\put(645,419){\usebox{\plotpoint}}
\put(646,418){\usebox{\plotpoint}}
\put(647,417){\usebox{\plotpoint}}
\put(649,414){\usebox{\plotpoint}}
\put(650,413){\usebox{\plotpoint}}
\put(651,412){\usebox{\plotpoint}}
\put(652,411){\usebox{\plotpoint}}
\put(653,409){\usebox{\plotpoint}}
\put(654,408){\usebox{\plotpoint}}
\put(655,407){\usebox{\plotpoint}}
\put(656,406){\usebox{\plotpoint}}
\put(657,404){\usebox{\plotpoint}}
\put(658,403){\usebox{\plotpoint}}
\put(659,402){\usebox{\plotpoint}}
\put(660,401){\usebox{\plotpoint}}
\put(661,399){\usebox{\plotpoint}}
\put(662,398){\usebox{\plotpoint}}
\put(663,397){\usebox{\plotpoint}}
\put(664,396){\usebox{\plotpoint}}
\put(665,394){\usebox{\plotpoint}}
\put(666,393){\usebox{\plotpoint}}
\put(667,392){\usebox{\plotpoint}}
\put(668,391){\usebox{\plotpoint}}
\put(669,389){\usebox{\plotpoint}}
\put(670,388){\usebox{\plotpoint}}
\put(671,387){\usebox{\plotpoint}}
\put(672,386){\usebox{\plotpoint}}
\put(673,384){\usebox{\plotpoint}}
\put(674,383){\usebox{\plotpoint}}
\put(675,382){\usebox{\plotpoint}}
\put(676,381){\usebox{\plotpoint}}
\put(677,379){\usebox{\plotpoint}}
\put(678,378){\usebox{\plotpoint}}
\put(679,377){\usebox{\plotpoint}}
\put(680,376){\usebox{\plotpoint}}
\put(681,375){\usebox{\plotpoint}}
\put(682,373){\usebox{\plotpoint}}
\put(683,372){\usebox{\plotpoint}}
\put(684,370){\usebox{\plotpoint}}
\put(685,369){\usebox{\plotpoint}}
\put(686,368){\usebox{\plotpoint}}
\put(687,366){\usebox{\plotpoint}}
\put(688,365){\usebox{\plotpoint}}
\put(689,363){\usebox{\plotpoint}}
\put(690,362){\usebox{\plotpoint}}
\put(691,361){\usebox{\plotpoint}}
\put(692,359){\usebox{\plotpoint}}
\put(693,358){\usebox{\plotpoint}}
\put(694,356){\usebox{\plotpoint}}
\put(695,355){\usebox{\plotpoint}}
\put(696,354){\usebox{\plotpoint}}
\put(697,352){\usebox{\plotpoint}}
\put(698,351){\usebox{\plotpoint}}
\put(699,349){\usebox{\plotpoint}}
\put(700,348){\usebox{\plotpoint}}
\put(701,347){\usebox{\plotpoint}}
\put(702,345){\usebox{\plotpoint}}
\put(703,344){\usebox{\plotpoint}}
\put(704,342){\usebox{\plotpoint}}
\put(705,341){\usebox{\plotpoint}}
\put(706,340){\usebox{\plotpoint}}
\put(707,338){\usebox{\plotpoint}}
\put(708,337){\usebox{\plotpoint}}
\put(709,335){\usebox{\plotpoint}}
\put(710,334){\usebox{\plotpoint}}
\put(711,333){\usebox{\plotpoint}}
\put(712,331){\usebox{\plotpoint}}
\put(713,330){\usebox{\plotpoint}}
\put(714,328){\usebox{\plotpoint}}
\put(715,327){\usebox{\plotpoint}}
\put(716,326){\usebox{\plotpoint}}
\put(717,324){\usebox{\plotpoint}}
\put(718,323){\usebox{\plotpoint}}
\put(719,322){\usebox{\plotpoint}}
\put(720,321){\usebox{\plotpoint}}
\put(721,319){\usebox{\plotpoint}}
\put(722,318){\usebox{\plotpoint}}
\put(723,317){\usebox{\plotpoint}}
\put(724,316){\usebox{\plotpoint}}
\put(725,314){\usebox{\plotpoint}}
\put(726,313){\usebox{\plotpoint}}
\put(727,312){\usebox{\plotpoint}}
\put(728,311){\usebox{\plotpoint}}
\put(729,309){\usebox{\plotpoint}}
\put(730,308){\usebox{\plotpoint}}
\put(731,307){\usebox{\plotpoint}}
\put(732,306){\usebox{\plotpoint}}
\put(733,304){\usebox{\plotpoint}}
\put(734,303){\usebox{\plotpoint}}
\put(735,302){\usebox{\plotpoint}}
\put(736,301){\usebox{\plotpoint}}
\put(737,300){\usebox{\plotpoint}}
\put(738,298){\usebox{\plotpoint}}
\put(739,297){\usebox{\plotpoint}}
\put(740,296){\usebox{\plotpoint}}
\put(741,295){\usebox{\plotpoint}}
\put(742,293){\usebox{\plotpoint}}
\put(743,292){\usebox{\plotpoint}}
\put(744,291){\usebox{\plotpoint}}
\put(745,290){\usebox{\plotpoint}}
\put(746,288){\usebox{\plotpoint}}
\put(747,287){\usebox{\plotpoint}}
\put(748,286){\usebox{\plotpoint}}
\put(749,285){\usebox{\plotpoint}}
\put(750,283){\usebox{\plotpoint}}
\put(751,282){\usebox{\plotpoint}}
\put(752,281){\usebox{\plotpoint}}
\put(753,280){\usebox{\plotpoint}}
\put(754,279){\usebox{\plotpoint}}
\put(755,279){\usebox{\plotpoint}}
\put(756,278){\usebox{\plotpoint}}
\put(757,277){\usebox{\plotpoint}}
\put(759,276){\usebox{\plotpoint}}
\put(760,275){\usebox{\plotpoint}}
\put(762,274){\usebox{\plotpoint}}
\put(763,273){\usebox{\plotpoint}}
\put(764,272){\usebox{\plotpoint}}
\put(766,271){\usebox{\plotpoint}}
\put(767,270){\usebox{\plotpoint}}
\put(769,269){\usebox{\plotpoint}}
\put(770,268){\usebox{\plotpoint}}
\put(771,267){\usebox{\plotpoint}}
\put(773,266){\usebox{\plotpoint}}
\put(774,265){\usebox{\plotpoint}}
\put(776,264){\usebox{\plotpoint}}
\put(777,263){\usebox{\plotpoint}}
\put(778,262){\usebox{\plotpoint}}
\put(780,261){\usebox{\plotpoint}}
\put(781,260){\usebox{\plotpoint}}
\put(783,259){\usebox{\plotpoint}}
\put(784,258){\usebox{\plotpoint}}
\put(785,257){\usebox{\plotpoint}}
\put(787,256){\usebox{\plotpoint}}
\put(788,255){\usebox{\plotpoint}}
\put(790,254){\usebox{\plotpoint}}
\put(791,253){\usebox{\plotpoint}}
\put(793,252){\usebox{\plotpoint}}
\put(795,251){\usebox{\plotpoint}}
\put(797,250){\usebox{\plotpoint}}
\put(799,249){\usebox{\plotpoint}}
\put(801,248){\usebox{\plotpoint}}
\put(803,247){\usebox{\plotpoint}}
\put(805,246){\usebox{\plotpoint}}
\put(807,245){\usebox{\plotpoint}}
\put(809,244){\usebox{\plotpoint}}
\put(811,243){\usebox{\plotpoint}}
\put(813,242){\usebox{\plotpoint}}
\put(815,241){\usebox{\plotpoint}}
\put(817,240){\usebox{\plotpoint}}
\put(819,239){\usebox{\plotpoint}}
\put(821,238){\usebox{\plotpoint}}
\put(823,237){\usebox{\plotpoint}}
\put(825,236){\usebox{\plotpoint}}
\put(827,235){\usebox{\plotpoint}}
\put(829,234){\usebox{\plotpoint}}
\put(831,233){\usebox{\plotpoint}}
\put(833,232){\usebox{\plotpoint}}
\put(836,231){\usebox{\plotpoint}}
\put(840,230){\usebox{\plotpoint}}
\put(844,229){\usebox{\plotpoint}}
\put(848,228){\usebox{\plotpoint}}
\put(852,227){\usebox{\plotpoint}}
\put(855,226){\usebox{\plotpoint}}
\put(859,225){\usebox{\plotpoint}}
\put(863,224){\usebox{\plotpoint}}
\put(867,223){\usebox{\plotpoint}}
\put(871,222){\usebox{\plotpoint}}
\put(874,221){\rule[-0.500pt]{1.514pt}{1.000pt}}
\put(881,220){\rule[-0.500pt]{1.514pt}{1.000pt}}
\put(887,219){\rule[-0.500pt]{1.514pt}{1.000pt}}
\put(893,218){\rule[-0.500pt]{1.514pt}{1.000pt}}
\put(900,217){\rule[-0.500pt]{1.514pt}{1.000pt}}
\put(906,216){\rule[-0.500pt]{1.514pt}{1.000pt}}
\put(912,215){\rule[-0.500pt]{1.514pt}{1.000pt}}
\put(499,565){\circle{24}}
\put(526,489){\circle{24}}
\put(555,468){\circle{24}}
\put(585,460){\circle{24}}
\put(617,445){\circle{24}}
\put(649,416){\circle{24}}
\put(682,375){\circle{24}}
\put(717,326){\circle{24}}
\put(755,279){\circle{24}}
\put(793,252){\circle{24}}
\put(833,232){\circle{24}}
\put(875,221){\circle{24}}
\put(919,214){\circle{24}}
\put(918,214){\usebox{\plotpoint}}
\put(330,391){\usebox{\plotpoint}}
\put(330,385){\rule[-0.500pt]{1.000pt}{1.317pt}}
\put(331,380){\rule[-0.500pt]{1.000pt}{1.317pt}}
\put(332,374){\rule[-0.500pt]{1.000pt}{1.317pt}}
\put(333,369){\rule[-0.500pt]{1.000pt}{1.317pt}}
\put(334,363){\rule[-0.500pt]{1.000pt}{1.317pt}}
\put(335,358){\rule[-0.500pt]{1.000pt}{1.317pt}}
\put(336,352){\rule[-0.500pt]{1.000pt}{1.317pt}}
\put(337,347){\rule[-0.500pt]{1.000pt}{1.317pt}}
\put(338,341){\rule[-0.500pt]{1.000pt}{1.317pt}}
\put(339,336){\rule[-0.500pt]{1.000pt}{1.317pt}}
\put(340,330){\rule[-0.500pt]{1.000pt}{1.317pt}}
\put(341,325){\rule[-0.500pt]{1.000pt}{1.317pt}}
\put(342,319){\rule[-0.500pt]{1.000pt}{1.317pt}}
\put(343,314){\rule[-0.500pt]{1.000pt}{1.317pt}}
\put(344,308){\rule[-0.500pt]{1.000pt}{1.317pt}}
\put(345,303){\rule[-0.500pt]{1.000pt}{1.317pt}}
\put(346,298){\rule[-0.500pt]{1.000pt}{1.317pt}}
\put(347,292){\rule[-0.500pt]{1.000pt}{1.317pt}}
\put(348,287){\rule[-0.500pt]{1.000pt}{1.317pt}}
\put(349,281){\rule[-0.500pt]{1.000pt}{1.317pt}}
\put(350,276){\rule[-0.500pt]{1.000pt}{1.317pt}}
\put(351,270){\rule[-0.500pt]{1.000pt}{1.317pt}}
\put(352,265){\rule[-0.500pt]{1.000pt}{1.317pt}}
\put(353,259){\rule[-0.500pt]{1.000pt}{1.317pt}}
\put(354,254){\rule[-0.500pt]{1.000pt}{1.317pt}}
\put(355,248){\rule[-0.500pt]{1.000pt}{1.317pt}}
\put(356,243){\rule[-0.500pt]{1.000pt}{1.317pt}}
\put(357,237){\rule[-0.500pt]{1.000pt}{1.317pt}}
\put(358,232){\rule[-0.500pt]{1.000pt}{1.317pt}}
\put(359,226){\rule[-0.500pt]{1.000pt}{1.317pt}}
\put(360,221){\rule[-0.500pt]{1.000pt}{1.317pt}}
\put(361,216){\rule[-0.500pt]{1.000pt}{1.317pt}}
\put(362,216){\usebox{\plotpoint}}
\put(364,215){\usebox{\plotpoint}}
\put(367,214){\usebox{\plotpoint}}
\put(370,213){\usebox{\plotpoint}}
\put(373,212){\usebox{\plotpoint}}
\put(376,211){\usebox{\plotpoint}}
\put(379,210){\usebox{\plotpoint}}
\put(382,209){\usebox{\plotpoint}}
\put(385,208){\usebox{\plotpoint}}
\put(387,207){\usebox{\plotpoint}}
\put(390,206){\usebox{\plotpoint}}
\put(393,205){\usebox{\plotpoint}}
\put(396,204){\usebox{\plotpoint}}
\put(399,203){\usebox{\plotpoint}}
\put(402,202){\usebox{\plotpoint}}
\put(405,201){\usebox{\plotpoint}}
\put(408,200){\usebox{\plotpoint}}
\put(411,201){\usebox{\plotpoint}}
\put(415,202){\usebox{\plotpoint}}
\put(419,203){\usebox{\plotpoint}}
\put(423,204){\usebox{\plotpoint}}
\put(426,205){\usebox{\plotpoint}}
\put(430,206){\usebox{\plotpoint}}
\put(434,207){\usebox{\plotpoint}}
\put(438,208){\usebox{\plotpoint}}
\put(442,209){\usebox{\plotpoint}}
\put(445,210){\usebox{\plotpoint}}
\put(449,211){\usebox{\plotpoint}}
\put(453,212){\usebox{\plotpoint}}
\put(457,213){\usebox{\plotpoint}}
\put(460,214){\rule[-0.500pt]{15.659pt}{1.000pt}}
\put(526,213){\rule[-0.500pt]{31.558pt}{1.000pt}}
\put(657,214){\rule[-0.500pt]{31.558pt}{1.000pt}}
\put(788,215){\rule[-0.500pt]{15.779pt}{1.000pt}}
\put(853,214){\rule[-0.500pt]{15.779pt}{1.000pt}}
\put(330,391){\circle*{24}}
\put(362,216){\circle*{24}}
\put(408,200){\circle*{24}}
\put(461,214){\circle*{24}}
\put(526,213){\circle*{24}}
\put(657,214){\circle*{24}}
\put(788,215){\circle*{24}}
\put(919,213){\circle*{24}}
\put(919,213){\rule[-0.500pt]{15.899pt}{1.000pt}}
\sbox{\plotpoint}{\rule[-0.350pt]{0.700pt}{0.700pt}}%
\put(330,410){\usebox{\plotpoint}}
\put(330,403){\rule[-0.350pt]{0.700pt}{1.641pt}}
\put(331,396){\rule[-0.350pt]{0.700pt}{1.641pt}}
\put(332,389){\rule[-0.350pt]{0.700pt}{1.641pt}}
\put(333,382){\rule[-0.350pt]{0.700pt}{1.641pt}}
\put(334,375){\rule[-0.350pt]{0.700pt}{1.641pt}}
\put(335,369){\rule[-0.350pt]{0.700pt}{1.641pt}}
\put(336,362){\rule[-0.350pt]{0.700pt}{1.641pt}}
\put(337,355){\rule[-0.350pt]{0.700pt}{1.641pt}}
\put(338,348){\rule[-0.350pt]{0.700pt}{1.641pt}}
\put(339,341){\rule[-0.350pt]{0.700pt}{1.641pt}}
\put(340,335){\rule[-0.350pt]{0.700pt}{1.641pt}}
\put(341,328){\rule[-0.350pt]{0.700pt}{1.641pt}}
\put(342,321){\rule[-0.350pt]{0.700pt}{1.641pt}}
\put(343,314){\rule[-0.350pt]{0.700pt}{1.641pt}}
\put(344,307){\rule[-0.350pt]{0.700pt}{1.641pt}}
\put(345,301){\rule[-0.350pt]{0.700pt}{1.641pt}}
\put(346,294){\rule[-0.350pt]{0.700pt}{1.641pt}}
\put(347,287){\rule[-0.350pt]{0.700pt}{1.641pt}}
\put(348,280){\rule[-0.350pt]{0.700pt}{1.641pt}}
\put(349,273){\rule[-0.350pt]{0.700pt}{1.641pt}}
\put(350,266){\rule[-0.350pt]{0.700pt}{1.641pt}}
\put(351,260){\rule[-0.350pt]{0.700pt}{1.641pt}}
\put(352,253){\rule[-0.350pt]{0.700pt}{1.641pt}}
\put(353,246){\rule[-0.350pt]{0.700pt}{1.641pt}}
\put(354,239){\rule[-0.350pt]{0.700pt}{1.641pt}}
\put(355,232){\rule[-0.350pt]{0.700pt}{1.641pt}}
\put(356,226){\rule[-0.350pt]{0.700pt}{1.641pt}}
\put(357,219){\rule[-0.350pt]{0.700pt}{1.641pt}}
\put(358,212){\rule[-0.350pt]{0.700pt}{1.641pt}}
\put(359,205){\rule[-0.350pt]{0.700pt}{1.641pt}}
\put(360,198){\rule[-0.350pt]{0.700pt}{1.641pt}}
\put(361,192){\rule[-0.350pt]{0.700pt}{1.641pt}}
\put(362,192){\usebox{\plotpoint}}
\put(363,193){\usebox{\plotpoint}}
\put(364,194){\usebox{\plotpoint}}
\put(365,195){\usebox{\plotpoint}}
\put(367,196){\usebox{\plotpoint}}
\put(368,197){\usebox{\plotpoint}}
\put(369,198){\usebox{\plotpoint}}
\put(370,199){\usebox{\plotpoint}}
\put(372,200){\usebox{\plotpoint}}
\put(373,201){\usebox{\plotpoint}}
\put(374,202){\usebox{\plotpoint}}
\put(376,203){\usebox{\plotpoint}}
\put(377,204){\usebox{\plotpoint}}
\put(378,205){\usebox{\plotpoint}}
\put(379,206){\usebox{\plotpoint}}
\put(381,207){\usebox{\plotpoint}}
\put(382,208){\usebox{\plotpoint}}
\put(383,209){\usebox{\plotpoint}}
\put(384,210){\usebox{\plotpoint}}
\put(386,211){\usebox{\plotpoint}}
\put(387,212){\usebox{\plotpoint}}
\put(388,213){\usebox{\plotpoint}}
\put(390,214){\usebox{\plotpoint}}
\put(391,215){\usebox{\plotpoint}}
\put(392,216){\usebox{\plotpoint}}
\put(393,217){\usebox{\plotpoint}}
\put(395,218){\usebox{\plotpoint}}
\put(396,219){\usebox{\plotpoint}}
\put(397,220){\usebox{\plotpoint}}
\put(399,221){\usebox{\plotpoint}}
\put(400,222){\usebox{\plotpoint}}
\put(401,223){\usebox{\plotpoint}}
\put(402,224){\usebox{\plotpoint}}
\put(404,225){\usebox{\plotpoint}}
\put(405,226){\usebox{\plotpoint}}
\put(406,227){\usebox{\plotpoint}}
\put(407,228){\rule[-0.350pt]{0.982pt}{0.700pt}}
\put(412,227){\rule[-0.350pt]{0.982pt}{0.700pt}}
\put(416,226){\rule[-0.350pt]{0.982pt}{0.700pt}}
\put(420,225){\rule[-0.350pt]{0.982pt}{0.700pt}}
\put(424,224){\rule[-0.350pt]{0.982pt}{0.700pt}}
\put(428,223){\rule[-0.350pt]{0.982pt}{0.700pt}}
\put(432,222){\rule[-0.350pt]{0.982pt}{0.700pt}}
\put(436,221){\rule[-0.350pt]{0.982pt}{0.700pt}}
\put(440,220){\rule[-0.350pt]{0.982pt}{0.700pt}}
\put(444,219){\rule[-0.350pt]{0.982pt}{0.700pt}}
\put(448,218){\rule[-0.350pt]{0.982pt}{0.700pt}}
\put(452,217){\rule[-0.350pt]{0.982pt}{0.700pt}}
\put(456,216){\rule[-0.350pt]{0.982pt}{0.700pt}}
\put(461,215){\rule[-0.350pt]{5.219pt}{0.700pt}}
\put(482,214){\rule[-0.350pt]{5.219pt}{0.700pt}}
\put(504,213){\rule[-0.350pt]{5.220pt}{0.700pt}}
\put(526,212){\rule[-0.350pt]{15.779pt}{0.700pt}}
\put(591,213){\rule[-0.350pt]{15.779pt}{0.700pt}}
\put(330,410){\raisebox{-1.2pt}{\makebox(0,0){$\Diamond$}}}
\put(362,192){\raisebox{-1.2pt}{\makebox(0,0){$\Diamond$}}}
\put(408,228){\raisebox{-1.2pt}{\makebox(0,0){$\Diamond$}}}
\put(461,215){\raisebox{-1.2pt}{\makebox(0,0){$\Diamond$}}}
\put(526,212){\raisebox{-1.2pt}{\makebox(0,0){$\Diamond$}}}
\put(657,214){\raisebox{-1.2pt}{\makebox(0,0){$\Diamond$}}}
\put(788,214){\raisebox{-1.2pt}{\makebox(0,0){$\Diamond$}}}
\put(919,214){\raisebox{-1.2pt}{\makebox(0,0){$\Diamond$}}}
\put(657,214){\rule[-0.350pt]{79.015pt}{0.700pt}}
\end{picture}
\vspace{-.1cm} }
\caption{Convergence of the total energy per carbon atom
vs grid fineness (given by the
cutoff $E_{cut}$ of the plane waves that it can represent).
The results of the present method are shown 
for a diamond supercell with 64 atoms (full circles)
and for a C$_3$ cluster (diamonds). Open circles show
results of conventional plane-wave calculations for diamond.$^{12}$}
\label{fig:ecut}
\end{figure}
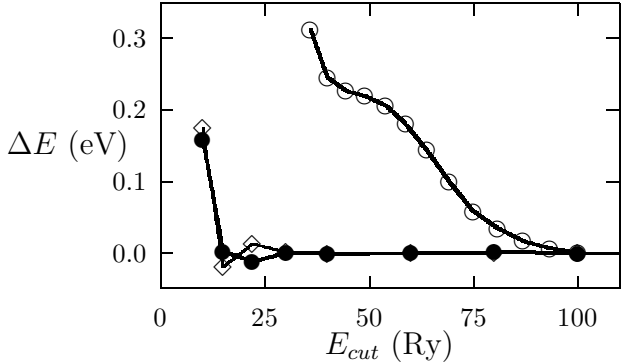

The LDA Hamiltonian matrix elements for a given particle density are
obtained using a combination of techniques, adopting the most convenient
one for each term of the Hamiltonian.
In a prior step, to avoid dealing with the long range of the pseudopotentials,
we rewrite the Kohn-Sham Hamiltonian by adding and 
subtracting the Hartree potential created by the neutral-atom charge 
$n_0({\bf r})$, defined as 
\begin{equation}
n_0({\bf r})=\sum_i n_i^{\rm NA} ({\bf r-R}_i) 
\label{atomcharge}
\end{equation}
where $i$ runs over the atoms in the system, and
$n_i^{\rm NA}$ is the spherical atomic
charge density of the atom $i$ in its neutral, isolated state
with $\rho^0_\mu$ electrons on each orbital $\phi_{\mu}$.
If we define $\delta n({\bf r})=n({\bf r})-n_0({\bf r})$ 
where $n({\bf r})$  is the actual charge density,
the Hartree potential can be decomposed into two contributions
$V_{\rm H}^\delta$ and $V_{\rm H}^0$, 
created by $\delta n({\bf r})$ and $n_0({\bf r})$
respectively. Using Eq. (\ref{atomcharge}),
$V_{\rm H}^0$  can be expressed as a sum
of atomic contributions.
Also, the pseudopotential is decomposed
into a short-range non-local term $V_{\rm NL}$
and a long-range local term $V_{\rm L}$.
Following Sankey and Niklewski,\cite{sankey} we define the
neutral atom potential of a given atom at ${\bf R}_i$
as
\begin{equation}
V_{\rm NA} ({\bf r-R}_i) = V_{\rm L} ({\bf r-R}_i) +
e^2 \int {n_i^{\rm NA} ({\bf r-R}_i)
\over |{\bf r-r}'|} d {\bf r}'.
\end{equation}
$V_{\rm NA}$ is short ranged, since the core
attraction and the electron Coulomb repulsion of the neutral
atom charge cancel each
other beyond $r_c$. 
The Kohn-Sham Hamiltonian is finally obtained as
\begin{eqnarray}
H^{\rm KS} &=& {p^2 \over 2m} 
+\sum_i \left[V_{\rm NL} ({\bf r-R}_i) 
+ V_{\rm NA} ({\bf r-R}_i)\right] \nonumber \\
&+& V_{\rm H}^\delta ({\bf r}) + V_{\rm XC}({\bf r}) 
\end{eqnarray}

The overlap, kinetic energy term, neutral
atom potential and non-local part of the pseudopotential,
are all independent of the charge density $n({\bf r})$,
and their matrix elements between atomic orbitals
can be expressed as sums of two center
($S_{\mu \nu}$=$\langle \phi_\mu | \phi_\nu \rangle$ and
$\langle \phi_\mu | p^2/2m | \phi_\nu \rangle$)
or three center
($\langle \phi_\mu | V_{\rm NL} ({\bf r-R}_i) | \phi_\nu \rangle$
and 
$\langle \phi_\mu | V_{\rm NA} ({\bf r-R}_i) | \phi_\nu \rangle$)
integrals, which only depend on the relative positions
of pairs or triplets of atoms. 
We follow the method proposed by Sankey and Niklewski\cite{sankey}
to compute all these integrals: they are calculated beforehand
and tabulated
as a function of the relative position of the centers.
These tables are used during the simulation, to calculate
all the non-zero integrals by interpolation. The details of the
procedure can be found in Ref.~\onlinecite{sankey}.
Since all these integrals are zero for distant enough
atoms, their number scales linearly with the size of the system,
as well as the computation time. The contributions of these terms
to the Hamiltonian are computed only once for a given atomic configuration,
since they do not depend on the selfconsistent charge.

The matrix elements of the Hartree potential $V_{\rm H}^\delta ({\bf r})$ 
created by the charge $\delta n({\bf r})$ and the exchange-correlation 
potential $V_{\rm XC} \left[ n({\bf r}) \right]$ both depend on the 
selfconsistent charge. 
To calculate these integrals we compute  $n_0({\bf r})$, $n({\bf r})$ and
$\delta n({\bf r})$ for a given LCAO density matrix at the points of a 
regular grid in real space. This is straightforward since the basis orbitals 
are defined in real space. 
Poisson's equation for the Hartree potential can be then solved
by the standard fast Fourier transform (FFT) method, assuming a
supercell geometry, or by the multigrid method.\cite{recipes}
In spite of its $N\log N$ scaling, we presently use FFT's for 
simplicity, since this part represents a minor contribution 
to the total computational load. Note that only two FFT's are 
necessary per cycle of selfconsistency (SCF cycle),
in contrast with PW based 
calculations, where an FFT is required for each electronic state. 
The LDA XC potential is trivially computed on each point of
the grid. Once the value of the Hartree and the XC potentials are known at
every point, the integrals 
$\langle \phi_\mu | V^\delta_H | \phi_\nu \rangle$ and
$\langle \phi_\mu | V_{XC} | \phi_\nu \rangle$ 
are computed by direct summation on the grid.
These sums are carefully done to minimize the amount of numerical
workload involved. Only the non-zero integrals (between orbitals 
on atoms closer than $2r_c$) are computed,
and only the points of the grid for which both
orbitals are non-zero contribute to each integral.
We use sparse-matrix multiplication techniques optimized
for this class of operations. As a result, the computation
of the integrals scales linearly with the size of the system.

It is important to stress that the convergence with grid spacing of
our method is different from that in standard PW
calculations, which are
known to require large PW cutoffs for systems containing
atoms with hard pseudopotentials. 
In Figure \ref{fig:ecut} we show the convergence of the 
total energy per atom (referred to the converged value) 
for carbon, as a function of 
$E_{cut}$, the kinetic energy cutoff of the plane waves
that the grid can represent.
Full circles are
for a diamond supercell of 64 atoms, 
whereas diamonds
are for a cluster of 3 carbon atoms in a supercell
of $15\times 15\times 15$ \AA$^3$.
In both cases, the results are converged to below 2 meV/atom
for a cutoff of 30 Ry. 
This is in sharp contrast with results of
PW calculations\cite{trullier} (open circles)
in which the cutoff necessary to 
achieve convergence (with the same pseudopotential)
is much higher. 
Note, moreover, that the energy
cutoff in our case refers to the grid representation of the
charge density, whereas in the PW case it refers to 
the wave functions, which implies an even higher 
(four times) cutoff in the charge density. The reason for the fast 
convergence of our approach is that most of the Hamiltonian terms
(most importantly the kinetic energy and the 
neutral atom potential)
are not computed in the grid.

Once the Kohn-Sham Hamiltonian has been obtained, 
we use a recently proposed
order-$N$ method\cite{ordejon1,kim}
to compute the band structure energy $E_{\rm BS}$ 
(sum of occupied eigenvalues).
In this approach, a modified band energy functional\cite{ordejon1,kim}
is minimized\cite{minimi} with respect to
the electronic orbitals by means
of a conjugate gradients (CG) algorithm, to yield $E_{\rm BS}$.
The orthonormality of the occupied states does not need
to be imposed, but it is obtained as a result of the minimization
of the energy functional. 
The elimination of the orthogonalization
is the first step to achieve an order-$N$ scaling. The second is
the use of localized, Wannier-like wave functions 
(LWF) to describe the
electronic states entering the minimization of the
energy functional. Truncation of these
localized functions beyond a given cutoff $R_c$ from
the center of the LWF provides a linear scaling
algorithm. The errors involved in this truncation,
which can be reduced arbitrarily by increasing the value of $R_c$,
are analyzed in detail in Ref~\onlinecite{ordejon}.

After the band energy has been minimized and the LWF's obtained,
the new charge density is computed, completing a so-called SCF cycle.
From the density, a new Hamiltonian is produced, the procedure being repeated 
until selfconsistency in the charge density or the Hamiltonian is achieved.
At this point, the total energy can be computed as
\begin{eqnarray}
E_{\rm tot} = &E&_{\rm BS} 
- {e^2 \over 2} \int V_{\rm H}({\bf r}) n({\bf r})  d{\bf r} 
+ {e^2 \over 2} \int V_{\rm H}^0({\bf r}) n_0({\bf r})  d{\bf r} 
\nonumber \\
&+& \int  \left[\epsilon_{\rm XC} (n) - V_{\rm XC}(n)
\right]  n({\bf r}) d{\bf r} +U_{\rm ii-ee}
\label{eq:enertot}
\end{eqnarray}
where $V_{\rm H}({\bf r})$ is the Hartree potential of the
selfconsistent charge $n({\bf r})$, and, following Sankey 
and Niklewski,\cite{sankey} we have defined
\begin{equation}
U_{\rm ii-ee} = {e^2 \over 2} {\sum_{l l'}}^\prime {Z_l Z_{l'} \over 
|{\bf R}_l-{\bf R}_{l'}|} - 
{e^2 \over 2} \int V_{\rm H}^0({\bf r}) n_0({\bf r})
\end{equation}
As in the case of the Hamiltonian, we have added and subtracted
the electrostatic energy of the neutral atom charge $n_0({\bf r})$ to
obtain Eq. (\ref{eq:enertot}). The advantage, again, is that $U_{\rm ii-ee}$
can be expressed as a sum of short range contributions,
which is easy to evaluate in $O(N)$ operations,\cite{sankey}
avoiding the problems related with the long range character
of the ionic core interactions. The integrals appearing in 
Eq. (\ref{eq:enertot}) are calculated in the real space grid.

\begin{figure}
\parbox{8.3cm}{ \vspace{-0.9cm} \hspace{-0.7cm}
%\input{figst}
% GNUPLOT: LaTeX picture
\setlength{\unitlength}{0.240900pt}
\ifx\plotpoint\undefined\newsavebox{\plotpoint}\fi
\sbox{\plotpoint}{\rule[-0.175pt]{0.350pt}{0.350pt}}%
\begin{picture}(840,689)(0,0)
\tenrm
\sbox{\plotpoint}{\rule[-0.175pt]{0.350pt}{0.350pt}}%
\put(264,230){\rule[-0.175pt]{4.818pt}{0.350pt}}
\put(242,230){\makebox(0,0)[r]{\normalsize $0$}}
\put(756,230){\rule[-0.175pt]{4.818pt}{0.350pt}}
\put(264,302){\rule[-0.175pt]{4.818pt}{0.350pt}}
\put(242,302){\makebox(0,0)[r]{\normalsize $50$}}
\put(756,302){\rule[-0.175pt]{4.818pt}{0.350pt}}
\put(264,374){\rule[-0.175pt]{4.818pt}{0.350pt}}
\put(242,374){\makebox(0,0)[r]{\normalsize $100$}}
\put(756,374){\rule[-0.175pt]{4.818pt}{0.350pt}}
\put(264,446){\rule[-0.175pt]{4.818pt}{0.350pt}}
\put(242,446){\makebox(0,0)[r]{\normalsize $150$}}
\put(756,446){\rule[-0.175pt]{4.818pt}{0.350pt}}
\put(264,518){\rule[-0.175pt]{4.818pt}{0.350pt}}
\put(242,518){\makebox(0,0)[r]{\normalsize $200$}}
\put(756,518){\rule[-0.175pt]{4.818pt}{0.350pt}}
\put(264,158){\rule[-0.175pt]{0.350pt}{4.818pt}}
\put(264,113){\makebox(0,0){\normalsize $0$}}
\put(264,556){\rule[-0.175pt]{0.350pt}{4.818pt}}
\put(371,158){\rule[-0.175pt]{0.350pt}{4.818pt}}
\put(371,113){\makebox(0,0){\normalsize $250$}}
\put(371,556){\rule[-0.175pt]{0.350pt}{4.818pt}}
\put(477,158){\rule[-0.175pt]{0.350pt}{4.818pt}}
\put(477,113){\makebox(0,0){\normalsize $500$}}
\put(477,556){\rule[-0.175pt]{0.350pt}{4.818pt}}
\put(584,158){\rule[-0.175pt]{0.350pt}{4.818pt}}
\put(584,113){\makebox(0,0){\normalsize $750$}}
\put(584,556){\rule[-0.175pt]{0.350pt}{4.818pt}}
\put(691,158){\rule[-0.175pt]{0.350pt}{4.818pt}}
\put(691,113){\makebox(0,0){\normalsize $1000$}}
\put(691,556){\rule[-0.175pt]{0.350pt}{4.818pt}}
\put(264,158){\rule[-0.175pt]{123.341pt}{0.350pt}}
\put(776,158){\rule[-0.175pt]{0.350pt}{100.696pt}}
\put(264,576){\rule[-0.175pt]{123.341pt}{0.350pt}}
\put(46,403){\makebox(0,0)[l]{\large SIZE}}
\put(55,338){\makebox(0,0)[l]{\normalsize (Mb)}}
\put(341,57){\makebox(0,0)[l]{\large Number of atoms}}
\put(891,417){\makebox(0,0)[l]{\large CPU}}
\put(891,360){\makebox(0,0)[l]{\large time}}
\put(895,302){\makebox(0,0)[l]{(min)}}
\put(789,158){\makebox(0,0)[l]{0}}
\put(789,230){\makebox(0,0)[l]{50}}
\put(789,302){\makebox(0,0)[l]{100}}
\put(789,374){\makebox(0,0)[l]{150}}
\put(789,446){\makebox(0,0)[l]{200}}
\put(789,518){\makebox(0,0)[l]{250}}
\put(264,158){\rule[-0.175pt]{0.350pt}{100.696pt}}
\put(618,288){\usebox{\plotpoint}}
\put(619,287){\usebox{\plotpoint}}
\put(620,286){\usebox{\plotpoint}}
\put(621,285){\usebox{\plotpoint}}
\put(622,284){\usebox{\plotpoint}}
\put(623,283){\usebox{\plotpoint}}
\put(624,282){\usebox{\plotpoint}}
\put(625,281){\usebox{\plotpoint}}
\put(626,280){\usebox{\plotpoint}}
\put(627,279){\usebox{\plotpoint}}
\put(628,278){\usebox{\plotpoint}}
\put(629,277){\usebox{\plotpoint}}
\put(630,276){\usebox{\plotpoint}}
\put(631,275){\usebox{\plotpoint}}
\put(632,274){\usebox{\plotpoint}}
\put(633,273){\usebox{\plotpoint}}
\put(634,272){\usebox{\plotpoint}}
\put(635,271){\usebox{\plotpoint}}
\put(636,270){\usebox{\plotpoint}}
\put(637,269){\usebox{\plotpoint}}
\put(638,268){\usebox{\plotpoint}}
\put(639,267){\usebox{\plotpoint}}
\put(640,266){\usebox{\plotpoint}}
\put(641,265){\usebox{\plotpoint}}
\put(642,264){\usebox{\plotpoint}}
\put(643,263){\usebox{\plotpoint}}
\put(644,262){\usebox{\plotpoint}}
\put(645,261){\usebox{\plotpoint}}
\put(648,259){\vector(1,0){64}}
\put(646,260){\usebox{\plotpoint}}
\put(441,432){\usebox{\plotpoint}}
\put(440,433){\usebox{\plotpoint}}
\put(439,434){\usebox{\plotpoint}}
\put(438,435){\usebox{\plotpoint}}
\put(437,436){\usebox{\plotpoint}}
\put(436,437){\usebox{\plotpoint}}
\put(435,438){\usebox{\plotpoint}}
\put(434,439){\usebox{\plotpoint}}
\put(433,440){\usebox{\plotpoint}}
\put(432,441){\usebox{\plotpoint}}
\put(431,442){\usebox{\plotpoint}}
\put(430,443){\usebox{\plotpoint}}
\put(429,444){\usebox{\plotpoint}}
\put(428,445){\usebox{\plotpoint}}
\put(427,446){\usebox{\plotpoint}}
\put(426,447){\usebox{\plotpoint}}
\put(425,448){\usebox{\plotpoint}}
\put(424,449){\usebox{\plotpoint}}
\put(423,450){\usebox{\plotpoint}}
\put(422,451){\usebox{\plotpoint}}
\put(421,452){\usebox{\plotpoint}}
\put(420,453){\usebox{\plotpoint}}
\put(419,454){\usebox{\plotpoint}}
\put(418,455){\usebox{\plotpoint}}
\put(417,456){\usebox{\plotpoint}}
\put(416,457){\usebox{\plotpoint}}
\put(415,458){\usebox{\plotpoint}}
\put(414,459){\usebox{\plotpoint}}
\put(413,461){\vector(-1,0){64}}
\put(413,460){\usebox{\plotpoint}}
\sbox{\plotpoint}{\rule[-0.350pt]{0.700pt}{0.700pt}}%
\put(291,165){\circle*{18}}
\put(356,209){\circle*{18}}
\put(482,285){\circle*{18}}
\put(691,408){\circle*{18}}
\put(282,158){\usebox{\plotpoint}}
\put(284,159){\usebox{\plotpoint}}
\put(285,160){\usebox{\plotpoint}}
\put(286,161){\usebox{\plotpoint}}
\put(287,162){\usebox{\plotpoint}}
\put(288,163){\rule[-0.350pt]{0.723pt}{0.700pt}}
\put(291,164){\usebox{\plotpoint}}
\put(292,165){\usebox{\plotpoint}}
\put(293,166){\usebox{\plotpoint}}
\put(295,167){\usebox{\plotpoint}}
\put(296,168){\usebox{\plotpoint}}
\put(297,169){\usebox{\plotpoint}}
\put(298,170){\usebox{\plotpoint}}
\put(300,171){\usebox{\plotpoint}}
\put(302,172){\usebox{\plotpoint}}
\put(303,173){\usebox{\plotpoint}}
\put(304,174){\usebox{\plotpoint}}
\put(305,175){\usebox{\plotpoint}}
\put(307,176){\usebox{\plotpoint}}
\put(309,177){\usebox{\plotpoint}}
\put(310,178){\usebox{\plotpoint}}
\put(311,179){\rule[-0.350pt]{0.723pt}{0.700pt}}
\put(314,180){\usebox{\plotpoint}}
\put(315,181){\usebox{\plotpoint}}
\put(316,182){\usebox{\plotpoint}}
\put(317,183){\usebox{\plotpoint}}
\put(318,184){\usebox{\plotpoint}}
\put(320,185){\usebox{\plotpoint}}
\put(321,186){\usebox{\plotpoint}}
\put(323,187){\usebox{\plotpoint}}
\put(325,188){\usebox{\plotpoint}}
\put(326,189){\usebox{\plotpoint}}
\put(327,190){\rule[-0.350pt]{0.723pt}{0.700pt}}
\put(330,191){\usebox{\plotpoint}}
\put(331,192){\usebox{\plotpoint}}
\put(332,193){\usebox{\plotpoint}}
\put(333,194){\usebox{\plotpoint}}
\put(334,195){\usebox{\plotpoint}}
\put(336,196){\usebox{\plotpoint}}
\put(337,197){\usebox{\plotpoint}}
\put(339,198){\usebox{\plotpoint}}
\put(341,199){\usebox{\plotpoint}}
\put(342,200){\usebox{\plotpoint}}
\put(343,201){\rule[-0.350pt]{0.723pt}{0.700pt}}
\put(346,202){\usebox{\plotpoint}}
\put(347,203){\usebox{\plotpoint}}
\put(348,204){\usebox{\plotpoint}}
\put(350,205){\usebox{\plotpoint}}
\put(351,206){\usebox{\plotpoint}}
\put(353,207){\usebox{\plotpoint}}
\put(355,208){\usebox{\plotpoint}}
\put(356,209){\usebox{\plotpoint}}
\put(357,210){\usebox{\plotpoint}}
\put(359,211){\usebox{\plotpoint}}
\put(360,212){\usebox{\plotpoint}}
\put(362,213){\usebox{\plotpoint}}
\put(364,214){\usebox{\plotpoint}}
\put(365,215){\usebox{\plotpoint}}
\put(366,216){\rule[-0.350pt]{0.723pt}{0.700pt}}
\put(369,217){\usebox{\plotpoint}}
\put(370,218){\usebox{\plotpoint}}
\put(371,219){\usebox{\plotpoint}}
\put(373,220){\usebox{\plotpoint}}
\put(374,221){\usebox{\plotpoint}}
\put(375,222){\rule[-0.350pt]{0.723pt}{0.700pt}}
\put(378,223){\usebox{\plotpoint}}
\put(379,224){\usebox{\plotpoint}}
\put(380,225){\usebox{\plotpoint}}
\put(382,226){\rule[-0.350pt]{0.723pt}{0.700pt}}
\put(385,227){\usebox{\plotpoint}}
\put(386,228){\usebox{\plotpoint}}
\put(387,229){\usebox{\plotpoint}}
\put(389,230){\usebox{\plotpoint}}
\put(390,231){\usebox{\plotpoint}}
\put(392,232){\usebox{\plotpoint}}
\put(394,233){\usebox{\plotpoint}}
\put(395,234){\usebox{\plotpoint}}
\put(396,235){\usebox{\plotpoint}}
\put(398,236){\rule[-0.350pt]{0.723pt}{0.700pt}}
\put(401,237){\usebox{\plotpoint}}
\put(402,238){\usebox{\plotpoint}}
\put(403,239){\usebox{\plotpoint}}
\put(405,240){\usebox{\plotpoint}}
\put(406,241){\usebox{\plotpoint}}
\put(408,242){\usebox{\plotpoint}}
\put(410,243){\usebox{\plotpoint}}
\put(412,244){\usebox{\plotpoint}}
\put(413,245){\usebox{\plotpoint}}
\put(414,246){\rule[-0.350pt]{0.723pt}{0.700pt}}
\put(417,247){\usebox{\plotpoint}}
\put(419,248){\usebox{\plotpoint}}
\put(420,249){\usebox{\plotpoint}}
\put(421,250){\rule[-0.350pt]{0.723pt}{0.700pt}}
\put(424,251){\usebox{\plotpoint}}
\put(426,252){\usebox{\plotpoint}}
\put(427,253){\usebox{\plotpoint}}
\put(428,254){\rule[-0.350pt]{0.723pt}{0.700pt}}
\put(431,255){\usebox{\plotpoint}}
\put(433,256){\usebox{\plotpoint}}
\put(434,257){\usebox{\plotpoint}}
\put(435,258){\usebox{\plotpoint}}
\put(437,259){\rule[-0.350pt]{0.723pt}{0.700pt}}
\put(440,260){\usebox{\plotpoint}}
\put(441,261){\usebox{\plotpoint}}
\put(442,262){\usebox{\plotpoint}}
\put(444,263){\rule[-0.350pt]{0.723pt}{0.700pt}}
\put(447,264){\usebox{\plotpoint}}
\put(448,265){\usebox{\plotpoint}}
\put(449,266){\usebox{\plotpoint}}
\put(451,267){\usebox{\plotpoint}}
\put(453,268){\usebox{\plotpoint}}
\put(454,269){\usebox{\plotpoint}}
\put(456,270){\usebox{\plotpoint}}
\put(458,271){\usebox{\plotpoint}}
\put(460,272){\usebox{\plotpoint}}
\put(461,273){\usebox{\plotpoint}}
\put(463,274){\usebox{\plotpoint}}
\put(465,275){\usebox{\plotpoint}}
\put(467,276){\usebox{\plotpoint}}
\put(468,277){\usebox{\plotpoint}}
\put(470,278){\usebox{\plotpoint}}
\put(472,279){\usebox{\plotpoint}}
\put(474,280){\usebox{\plotpoint}}
\put(475,281){\usebox{\plotpoint}}
\put(476,282){\rule[-0.350pt]{0.723pt}{0.700pt}}
\put(479,283){\usebox{\plotpoint}}
\put(481,284){\usebox{\plotpoint}}
\put(482,285){\usebox{\plotpoint}}
\put(483,286){\rule[-0.350pt]{0.723pt}{0.700pt}}
\put(486,287){\usebox{\plotpoint}}
\put(488,288){\usebox{\plotpoint}}
\put(489,289){\usebox{\plotpoint}}
\put(490,290){\usebox{\plotpoint}}
\put(492,291){\rule[-0.350pt]{0.723pt}{0.700pt}}
\put(495,292){\usebox{\plotpoint}}
\put(496,293){\usebox{\plotpoint}}
\put(497,294){\usebox{\plotpoint}}
\put(499,295){\rule[-0.350pt]{0.723pt}{0.700pt}}
\put(502,296){\usebox{\plotpoint}}
\put(503,297){\usebox{\plotpoint}}
\put(504,298){\usebox{\plotpoint}}
\put(506,299){\rule[-0.350pt]{0.723pt}{0.700pt}}
\put(509,300){\usebox{\plotpoint}}
\put(510,301){\usebox{\plotpoint}}
\put(511,302){\usebox{\plotpoint}}
\put(513,303){\usebox{\plotpoint}}
\put(515,304){\usebox{\plotpoint}}
\put(516,305){\usebox{\plotpoint}}
\put(518,306){\usebox{\plotpoint}}
\put(520,307){\usebox{\plotpoint}}
\put(522,308){\usebox{\plotpoint}}
\put(523,309){\usebox{\plotpoint}}
\put(525,310){\usebox{\plotpoint}}
\put(527,311){\usebox{\plotpoint}}
\put(529,312){\usebox{\plotpoint}}
\put(530,313){\usebox{\plotpoint}}
\put(531,314){\rule[-0.350pt]{0.723pt}{0.700pt}}
\put(534,315){\usebox{\plotpoint}}
\put(536,316){\usebox{\plotpoint}}
\put(537,317){\usebox{\plotpoint}}
\put(538,318){\rule[-0.350pt]{0.723pt}{0.700pt}}
\put(541,319){\usebox{\plotpoint}}
\put(543,320){\usebox{\plotpoint}}
\put(544,321){\usebox{\plotpoint}}
\put(545,322){\rule[-0.350pt]{0.723pt}{0.700pt}}
\put(548,323){\usebox{\plotpoint}}
\put(550,324){\usebox{\plotpoint}}
\put(551,325){\usebox{\plotpoint}}
\put(552,326){\usebox{\plotpoint}}
\put(554,327){\usebox{\plotpoint}}
\put(555,328){\usebox{\plotpoint}}
\put(557,329){\usebox{\plotpoint}}
\put(559,330){\usebox{\plotpoint}}
\put(561,331){\usebox{\plotpoint}}
\put(562,332){\usebox{\plotpoint}}
\put(564,333){\usebox{\plotpoint}}
\put(566,334){\usebox{\plotpoint}}
\put(568,335){\usebox{\plotpoint}}
\put(569,336){\usebox{\plotpoint}}
\put(570,337){\rule[-0.350pt]{0.723pt}{0.700pt}}
\put(573,338){\usebox{\plotpoint}}
\put(575,339){\usebox{\plotpoint}}
\put(576,340){\usebox{\plotpoint}}
\put(577,341){\rule[-0.350pt]{0.723pt}{0.700pt}}
\put(580,342){\usebox{\plotpoint}}
\put(582,343){\usebox{\plotpoint}}
\put(583,344){\usebox{\plotpoint}}
\put(584,345){\rule[-0.350pt]{0.723pt}{0.700pt}}
\put(587,346){\usebox{\plotpoint}}
\put(589,347){\usebox{\plotpoint}}
\put(590,348){\usebox{\plotpoint}}
\put(591,349){\usebox{\plotpoint}}
\put(593,350){\usebox{\plotpoint}}
\put(594,351){\usebox{\plotpoint}}
\put(596,352){\usebox{\plotpoint}}
\put(598,353){\usebox{\plotpoint}}
\put(600,354){\usebox{\plotpoint}}
\put(601,355){\usebox{\plotpoint}}
\put(603,356){\usebox{\plotpoint}}
\put(605,357){\usebox{\plotpoint}}
\put(607,358){\usebox{\plotpoint}}
\put(608,359){\usebox{\plotpoint}}
\put(609,360){\rule[-0.350pt]{0.723pt}{0.700pt}}
\put(612,361){\usebox{\plotpoint}}
\put(614,362){\usebox{\plotpoint}}
\put(615,363){\usebox{\plotpoint}}
\put(616,364){\rule[-0.350pt]{0.723pt}{0.700pt}}
\put(619,365){\usebox{\plotpoint}}
\put(621,366){\usebox{\plotpoint}}
\put(622,367){\usebox{\plotpoint}}
\put(623,368){\rule[-0.350pt]{0.723pt}{0.700pt}}
\put(626,369){\usebox{\plotpoint}}
\put(627,370){\usebox{\plotpoint}}
\put(628,371){\usebox{\plotpoint}}
\put(630,372){\usebox{\plotpoint}}
\put(632,373){\usebox{\plotpoint}}
\put(633,374){\usebox{\plotpoint}}
\put(635,375){\usebox{\plotpoint}}
\put(637,376){\usebox{\plotpoint}}
\put(639,377){\usebox{\plotpoint}}
\put(640,378){\usebox{\plotpoint}}
\put(642,379){\usebox{\plotpoint}}
\put(644,380){\usebox{\plotpoint}}
\put(646,381){\usebox{\plotpoint}}
\put(647,382){\usebox{\plotpoint}}
\put(648,383){\rule[-0.350pt]{0.723pt}{0.700pt}}
\put(651,384){\usebox{\plotpoint}}
\put(652,385){\usebox{\plotpoint}}
\put(653,386){\usebox{\plotpoint}}
\put(655,387){\rule[-0.350pt]{0.723pt}{0.700pt}}
\put(658,388){\usebox{\plotpoint}}
\put(659,389){\usebox{\plotpoint}}
\put(660,390){\usebox{\plotpoint}}
\put(662,391){\rule[-0.350pt]{0.723pt}{0.700pt}}
\put(665,392){\usebox{\plotpoint}}
\put(666,393){\usebox{\plotpoint}}
\put(667,394){\usebox{\plotpoint}}
\put(669,395){\usebox{\plotpoint}}
\put(671,396){\usebox{\plotpoint}}
\put(672,397){\usebox{\plotpoint}}
\put(674,398){\usebox{\plotpoint}}
\put(676,399){\usebox{\plotpoint}}
\put(677,400){\usebox{\plotpoint}}
\put(678,401){\rule[-0.350pt]{0.723pt}{0.700pt}}
\put(681,402){\usebox{\plotpoint}}
\put(683,403){\usebox{\plotpoint}}
\put(684,404){\usebox{\plotpoint}}
\put(685,405){\usebox{\plotpoint}}
\put(687,406){\rule[-0.350pt]{0.723pt}{0.700pt}}
\put(690,407){\usebox{\plotpoint}}
\put(691,408){\usebox{\plotpoint}}
\put(692,409){\usebox{\plotpoint}}
\put(694,410){\rule[-0.350pt]{0.723pt}{0.700pt}}
\put(697,411){\usebox{\plotpoint}}
\put(698,412){\usebox{\plotpoint}}
\put(699,413){\usebox{\plotpoint}}
\put(701,414){\usebox{\plotpoint}}
\put(702,415){\usebox{\plotpoint}}
\put(704,416){\usebox{\plotpoint}}
\put(706,417){\usebox{\plotpoint}}
\put(708,418){\usebox{\plotpoint}}
\put(709,419){\usebox{\plotpoint}}
\put(710,420){\rule[-0.350pt]{0.723pt}{0.700pt}}
\put(713,421){\usebox{\plotpoint}}
\put(715,422){\usebox{\plotpoint}}
\put(716,423){\usebox{\plotpoint}}
\put(717,424){\rule[-0.350pt]{0.723pt}{0.700pt}}
\put(720,425){\usebox{\plotpoint}}
\put(722,426){\usebox{\plotpoint}}
\put(723,427){\usebox{\plotpoint}}
\put(724,428){\usebox{\plotpoint}}
\put(726,429){\usebox{\plotpoint}}
\put(727,430){\usebox{\plotpoint}}
\put(729,431){\usebox{\plotpoint}}
\put(731,432){\usebox{\plotpoint}}
\put(291,252){\circle{18}}
\put(356,290){\circle{18}}
\put(482,375){\circle{18}}
\put(691,503){\circle{18}}
\put(277,244){\usebox{\plotpoint}}
\put(277,244){\usebox{\plotpoint}}
\put(279,245){\usebox{\plotpoint}}
\put(281,246){\usebox{\plotpoint}}
\put(282,247){\usebox{\plotpoint}}
\put(284,248){\usebox{\plotpoint}}
\put(286,249){\usebox{\plotpoint}}
\put(288,250){\usebox{\plotpoint}}
\put(289,251){\usebox{\plotpoint}}
\put(291,252){\usebox{\plotpoint}}
\put(293,253){\usebox{\plotpoint}}
\put(295,254){\usebox{\plotpoint}}
\put(297,255){\usebox{\plotpoint}}
\put(298,256){\usebox{\plotpoint}}
\put(300,257){\usebox{\plotpoint}}
\put(302,258){\usebox{\plotpoint}}
\put(304,259){\usebox{\plotpoint}}
\put(305,260){\usebox{\plotpoint}}
\put(307,261){\usebox{\plotpoint}}
\put(309,262){\usebox{\plotpoint}}
\put(311,263){\usebox{\plotpoint}}
\put(312,264){\usebox{\plotpoint}}
\put(314,265){\usebox{\plotpoint}}
\put(316,266){\usebox{\plotpoint}}
\put(318,267){\usebox{\plotpoint}}
\put(320,268){\usebox{\plotpoint}}
\put(321,269){\usebox{\plotpoint}}
\put(323,270){\usebox{\plotpoint}}
\put(325,271){\usebox{\plotpoint}}
\put(327,272){\usebox{\plotpoint}}
\put(328,273){\usebox{\plotpoint}}
\put(330,274){\usebox{\plotpoint}}
\put(332,275){\usebox{\plotpoint}}
\put(334,276){\usebox{\plotpoint}}
\put(335,277){\usebox{\plotpoint}}
\put(336,278){\rule[-0.350pt]{0.723pt}{0.700pt}}
\put(339,279){\usebox{\plotpoint}}
\put(340,280){\usebox{\plotpoint}}
\put(341,281){\usebox{\plotpoint}}
\put(343,282){\rule[-0.350pt]{0.723pt}{0.700pt}}
\put(346,283){\usebox{\plotpoint}}
\put(347,284){\usebox{\plotpoint}}
\put(348,285){\usebox{\plotpoint}}
\put(350,286){\rule[-0.350pt]{0.723pt}{0.700pt}}
\put(353,287){\usebox{\plotpoint}}
\put(354,288){\usebox{\plotpoint}}
\put(355,289){\usebox{\plotpoint}}
\put(357,290){\usebox{\plotpoint}}
\put(358,291){\usebox{\plotpoint}}
\put(359,292){\rule[-0.350pt]{0.723pt}{0.700pt}}
\put(362,293){\usebox{\plotpoint}}
\put(363,294){\usebox{\plotpoint}}
\put(364,295){\usebox{\plotpoint}}
\put(366,296){\usebox{\plotpoint}}
\put(367,297){\usebox{\plotpoint}}
\put(369,298){\usebox{\plotpoint}}
\put(371,299){\usebox{\plotpoint}}
\put(372,300){\usebox{\plotpoint}}
\put(373,301){\usebox{\plotpoint}}
\put(375,302){\rule[-0.350pt]{0.723pt}{0.700pt}}
\put(378,303){\usebox{\plotpoint}}
\put(379,304){\usebox{\plotpoint}}
\put(380,305){\usebox{\plotpoint}}
\put(382,306){\usebox{\plotpoint}}
\put(383,307){\usebox{\plotpoint}}
\put(385,308){\usebox{\plotpoint}}
\put(386,309){\usebox{\plotpoint}}
\put(387,310){\usebox{\plotpoint}}
\put(389,311){\usebox{\plotpoint}}
\put(390,312){\usebox{\plotpoint}}
\put(392,313){\usebox{\plotpoint}}
\put(394,314){\usebox{\plotpoint}}
\put(395,315){\usebox{\plotpoint}}
\put(396,316){\usebox{\plotpoint}}
\put(398,317){\usebox{\plotpoint}}
\put(399,318){\usebox{\plotpoint}}
\put(401,319){\usebox{\plotpoint}}
\put(403,320){\usebox{\plotpoint}}
\put(404,321){\usebox{\plotpoint}}
\put(405,322){\rule[-0.350pt]{0.723pt}{0.700pt}}
\put(408,323){\usebox{\plotpoint}}
\put(409,324){\usebox{\plotpoint}}
\put(410,325){\usebox{\plotpoint}}
\put(411,326){\usebox{\plotpoint}}
\put(412,327){\usebox{\plotpoint}}
\put(414,328){\usebox{\plotpoint}}
\put(415,329){\usebox{\plotpoint}}
\put(417,330){\usebox{\plotpoint}}
\put(419,331){\usebox{\plotpoint}}
\put(420,332){\usebox{\plotpoint}}
\put(421,333){\rule[-0.350pt]{0.723pt}{0.700pt}}
\put(424,334){\usebox{\plotpoint}}
\put(425,335){\usebox{\plotpoint}}
\put(426,336){\usebox{\plotpoint}}
\put(427,337){\usebox{\plotpoint}}
\put(428,338){\rule[-0.350pt]{0.723pt}{0.700pt}}
\put(431,339){\usebox{\plotpoint}}
\put(432,340){\usebox{\plotpoint}}
\put(433,341){\usebox{\plotpoint}}
\put(435,342){\usebox{\plotpoint}}
\put(436,343){\usebox{\plotpoint}}
\put(437,344){\usebox{\plotpoint}}
\put(438,345){\usebox{\plotpoint}}
\put(440,346){\usebox{\plotpoint}}
\put(442,347){\usebox{\plotpoint}}
\put(443,348){\usebox{\plotpoint}}
\put(444,349){\rule[-0.350pt]{0.723pt}{0.700pt}}
\put(447,350){\usebox{\plotpoint}}
\put(448,351){\usebox{\plotpoint}}
\put(449,352){\usebox{\plotpoint}}
\put(450,353){\usebox{\plotpoint}}
\put(451,354){\usebox{\plotpoint}}
\put(453,355){\usebox{\plotpoint}}
\put(454,356){\usebox{\plotpoint}}
\put(456,357){\usebox{\plotpoint}}
\put(458,358){\usebox{\plotpoint}}
\put(459,359){\usebox{\plotpoint}}
\put(460,360){\rule[-0.350pt]{0.723pt}{0.700pt}}
\put(463,361){\usebox{\plotpoint}}
\put(464,362){\usebox{\plotpoint}}
\put(465,363){\usebox{\plotpoint}}
\put(466,364){\usebox{\plotpoint}}
\put(467,365){\rule[-0.350pt]{0.723pt}{0.700pt}}
\put(470,366){\usebox{\plotpoint}}
\put(471,367){\usebox{\plotpoint}}
\put(472,368){\usebox{\plotpoint}}
\put(474,369){\usebox{\plotpoint}}
\put(475,370){\usebox{\plotpoint}}
\put(476,371){\rule[-0.350pt]{0.723pt}{0.700pt}}
\put(479,372){\usebox{\plotpoint}}
\put(480,373){\usebox{\plotpoint}}
\put(481,374){\usebox{\plotpoint}}
\put(483,375){\usebox{\plotpoint}}
\put(484,376){\usebox{\plotpoint}}
\put(486,377){\usebox{\plotpoint}}
\put(487,378){\usebox{\plotpoint}}
\put(488,379){\usebox{\plotpoint}}
\put(490,380){\usebox{\plotpoint}}
\put(491,381){\usebox{\plotpoint}}
\put(492,382){\rule[-0.350pt]{0.723pt}{0.700pt}}
\put(495,383){\usebox{\plotpoint}}
\put(496,384){\usebox{\plotpoint}}
\put(497,385){\usebox{\plotpoint}}
\put(499,386){\usebox{\plotpoint}}
\put(500,387){\usebox{\plotpoint}}
\put(502,388){\usebox{\plotpoint}}
\put(504,389){\usebox{\plotpoint}}
\put(505,390){\usebox{\plotpoint}}
\put(506,391){\rule[-0.350pt]{0.723pt}{0.700pt}}
\put(509,392){\usebox{\plotpoint}}
\put(510,393){\usebox{\plotpoint}}
\put(511,394){\usebox{\plotpoint}}
\put(513,395){\usebox{\plotpoint}}
\put(514,396){\usebox{\plotpoint}}
\put(515,397){\rule[-0.350pt]{0.723pt}{0.700pt}}
\put(518,398){\usebox{\plotpoint}}
\put(519,399){\usebox{\plotpoint}}
\put(520,400){\usebox{\plotpoint}}
\put(522,401){\usebox{\plotpoint}}
\put(523,402){\usebox{\plotpoint}}
\put(525,403){\usebox{\plotpoint}}
\put(527,404){\usebox{\plotpoint}}
\put(528,405){\usebox{\plotpoint}}
\put(529,406){\usebox{\plotpoint}}
\put(531,407){\rule[-0.350pt]{0.723pt}{0.700pt}}
\put(534,408){\usebox{\plotpoint}}
\put(535,409){\usebox{\plotpoint}}
\put(536,410){\usebox{\plotpoint}}
\put(538,411){\usebox{\plotpoint}}
\put(539,412){\usebox{\plotpoint}}
\put(541,413){\usebox{\plotpoint}}
\put(543,414){\usebox{\plotpoint}}
\put(544,415){\usebox{\plotpoint}}
\put(545,416){\rule[-0.350pt]{0.723pt}{0.700pt}}
\put(548,417){\usebox{\plotpoint}}
\put(549,418){\usebox{\plotpoint}}
\put(550,419){\usebox{\plotpoint}}
\put(552,420){\usebox{\plotpoint}}
\put(554,421){\usebox{\plotpoint}}
\put(555,422){\usebox{\plotpoint}}
\put(557,423){\usebox{\plotpoint}}
\put(559,424){\usebox{\plotpoint}}
\put(560,425){\usebox{\plotpoint}}
\put(561,426){\rule[-0.350pt]{0.723pt}{0.700pt}}
\put(564,427){\usebox{\plotpoint}}
\put(565,428){\usebox{\plotpoint}}
\put(566,429){\usebox{\plotpoint}}
\put(568,430){\usebox{\plotpoint}}
\put(570,431){\usebox{\plotpoint}}
\put(571,432){\usebox{\plotpoint}}
\put(573,433){\usebox{\plotpoint}}
\put(575,434){\usebox{\plotpoint}}
\put(576,435){\usebox{\plotpoint}}
\put(577,436){\rule[-0.350pt]{0.723pt}{0.700pt}}
\put(580,437){\usebox{\plotpoint}}
\put(582,438){\usebox{\plotpoint}}
\put(583,439){\usebox{\plotpoint}}
\put(584,440){\rule[-0.350pt]{0.723pt}{0.700pt}}
\put(587,441){\usebox{\plotpoint}}
\put(588,442){\usebox{\plotpoint}}
\put(589,443){\usebox{\plotpoint}}
\put(591,444){\usebox{\plotpoint}}
\put(593,445){\usebox{\plotpoint}}
\put(594,446){\usebox{\plotpoint}}
\put(596,447){\usebox{\plotpoint}}
\put(598,448){\usebox{\plotpoint}}
\put(600,449){\usebox{\plotpoint}}
\put(601,450){\usebox{\plotpoint}}
\put(603,451){\usebox{\plotpoint}}
\put(605,452){\usebox{\plotpoint}}
\put(606,453){\usebox{\plotpoint}}
\put(607,454){\usebox{\plotpoint}}
\put(609,455){\rule[-0.350pt]{0.723pt}{0.700pt}}
\put(612,456){\usebox{\plotpoint}}
\put(613,457){\usebox{\plotpoint}}
\put(614,458){\usebox{\plotpoint}}
\put(616,459){\rule[-0.350pt]{0.723pt}{0.700pt}}
\put(619,460){\usebox{\plotpoint}}
\put(620,461){\usebox{\plotpoint}}
\put(621,462){\usebox{\plotpoint}}
\put(623,463){\rule[-0.350pt]{0.723pt}{0.700pt}}
\put(626,464){\usebox{\plotpoint}}
\put(627,465){\usebox{\plotpoint}}
\put(628,466){\usebox{\plotpoint}}
\put(630,467){\usebox{\plotpoint}}
\put(631,468){\usebox{\plotpoint}}
\put(632,469){\rule[-0.350pt]{0.723pt}{0.700pt}}
\put(635,470){\usebox{\plotpoint}}
\put(637,471){\usebox{\plotpoint}}
\put(638,472){\usebox{\plotpoint}}
\put(639,473){\rule[-0.350pt]{0.723pt}{0.700pt}}
\put(642,474){\usebox{\plotpoint}}
\put(644,475){\usebox{\plotpoint}}
\put(645,476){\usebox{\plotpoint}}
\put(646,477){\usebox{\plotpoint}}
\put(648,478){\rule[-0.350pt]{0.723pt}{0.700pt}}
\put(651,479){\usebox{\plotpoint}}
\put(652,480){\usebox{\plotpoint}}
\put(653,481){\usebox{\plotpoint}}
\put(655,482){\rule[-0.350pt]{0.723pt}{0.700pt}}
\put(658,483){\usebox{\plotpoint}}
\put(659,484){\usebox{\plotpoint}}
\put(660,485){\usebox{\plotpoint}}
\put(662,486){\rule[-0.350pt]{0.723pt}{0.700pt}}
\put(665,487){\usebox{\plotpoint}}
\put(666,488){\usebox{\plotpoint}}
\put(667,489){\usebox{\plotpoint}}
\put(669,490){\usebox{\plotpoint}}
\put(671,491){\usebox{\plotpoint}}
\put(672,492){\usebox{\plotpoint}}
\put(674,493){\usebox{\plotpoint}}
\put(676,494){\usebox{\plotpoint}}
\put(678,495){\usebox{\plotpoint}}
\put(679,496){\usebox{\plotpoint}}
\put(681,497){\usebox{\plotpoint}}
\put(683,498){\usebox{\plotpoint}}
\put(685,499){\usebox{\plotpoint}}
\put(686,500){\usebox{\plotpoint}}
\put(687,501){\rule[-0.350pt]{0.723pt}{0.700pt}}
\put(690,502){\usebox{\plotpoint}}
\put(692,503){\usebox{\plotpoint}}
\put(693,504){\usebox{\plotpoint}}
\put(694,505){\rule[-0.350pt]{0.723pt}{0.700pt}}
\put(697,506){\usebox{\plotpoint}}
\put(699,507){\usebox{\plotpoint}}
\put(700,508){\usebox{\plotpoint}}
\put(701,509){\rule[-0.350pt]{0.723pt}{0.700pt}}
\put(704,510){\usebox{\plotpoint}}
\put(706,511){\usebox{\plotpoint}}
\put(707,512){\usebox{\plotpoint}}
\put(708,513){\usebox{\plotpoint}}
\put(710,514){\rule[-0.350pt]{0.723pt}{0.700pt}}
\put(713,515){\usebox{\plotpoint}}
\put(714,516){\usebox{\plotpoint}}
\put(715,517){\usebox{\plotpoint}}
\put(717,518){\rule[-0.350pt]{0.723pt}{0.700pt}}
\put(720,519){\usebox{\plotpoint}}
\put(721,520){\usebox{\plotpoint}}
\put(722,521){\usebox{\plotpoint}}
\put(724,522){\usebox{\plotpoint}}
\put(725,523){\usebox{\plotpoint}}
\put(726,524){\rule[-0.350pt]{0.723pt}{0.700pt}}
\put(729,525){\usebox{\plotpoint}}
\put(731,526){\usebox{\plotpoint}}
\put(732,527){\usebox{\plotpoint}}
\end{picture}
\vspace{-.1cm} }
\caption{
CPU time per SCF cycle and job memory for a simulation of 
Si supercells with different sizes. Times measured in an IBM
PowerPC with 17 Mflops (Linpack 100x100).}
\label{fig:Si}
\end{figure}
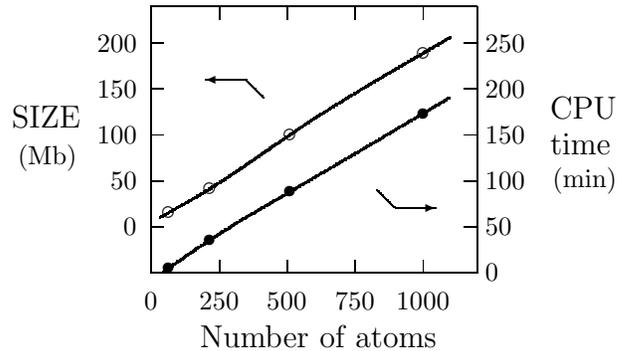

In molecular dynamics (MD) simulations and geometrical
optimizations the atomic forces are needed. We compute them
using a variation of the Hellman-Feynman theorem,
which includes Pulay-like corrections to account for the
fact that the basis set is not complete and moves
with the atoms. The force on atom $i$ is
\begin{eqnarray}
&&{\bf F}_i 
= - \sum_{\mu \nu} \rho_{\mu \nu} {{\partial} H_{\mu \nu}^0
\over \partial {\bf R}_i}
+ \sum_{\mu \nu} E_{\mu \nu} {{\partial} S_{\mu \nu}
\over \partial {\bf R}_i}  
- {\partial U_{\rm ii-ee} \over \partial {\bf R}_i} 
\nonumber \\
&+& \hspace{-0.06truecm} 2 \hspace{-0.06truecm} 
\sum_{\mu} \hspace{-0.06truecm} \rho_{\mu}^0 \langle 
{\partial \phi_\mu \over \partial {\bf R}_i} | 
V_{\rm H}^\delta |
\phi_\mu \hspace{-0.03truecm} \rangle \hspace{-0.03truecm}
- \hspace{-0.06truecm} 2 \hspace{-0.06truecm}  
\sum_{\mu \nu} \hspace{-0.06truecm} \rho_{\mu \nu} \langle 
{\partial \phi_\mu \over \partial {\bf R}_i} | 
V_{\rm H}^\delta + V_{\rm XC} |
\phi_\nu \rangle
\end{eqnarray} 
where $H^0 = p^2/2m + V_{\rm NL} + V_{\rm NA}$,
and $\rho_{\mu \nu}$ and $E_{\mu \nu}$ are the
density and energy-density matrices, respectively.\cite{sankey}
The first three terms are calculated interpolating 
the table data,\cite{sankey} whereas the last two
terms are computed by numerical integration
in the grid, as was done for the matrix elements
of the Hartree and XC potentials in the Kohn-Sham
Hamiltonian.

In order to show the linear scaling of the method, we have
performed calculations on supercells of silicon in the
diamond structure, with different numbers of atoms
from 64 to 1000. Only the $\Gamma$ point was used to 
sample the Brillouin zone, the cutoff for the charge density
grid was 12 Ry, and the LWF's were truncated at 4.5 \AA.
Figure \ref{fig:Si} shows the linear behavior of the
CPU time and memory requirements with the number of atoms.
The CPU time shown represents the average
cost to perform a SCF step in a  MD simulation, including
the calculation of the charge density and Hamiltonian matrix 
elements, the minimization of the band structure energy, 
and the calculation of the atomic forces.
The band structure energy minimization within each SCF cycle required
an average of 20 CG iterations, while the number of SCF cycles
depends largely on the simulation temperature, length of the time step and
mixing algorithm for selfconsistency. So far, in comparable
simulation conditions, no significant
dependence of the number of CG iterations and SCF cycles
with the size of the system has been observed. 
As we can see, in the present method both the CPU time and memory requirements 
are small enough to permit the calculation of a system of 1000 silicon
atoms in a very modest workstation.

\noindent
\parbox{8.3cm}{ \hspace{.0cm} \parbox{8.3cm}{ \vspace{-.4cm}
\begin{table}
\caption{
Average radius ($\bar{r}$), standard ($\sigma_s$) and 
maximum deviation [$\sigma_{m}=(r_{max}-r_{min})/2$]
of radii, and non-planarity angle$^3$ $\phi$
around pentagons (going from 0$^{\circ}$ for a planar pentagonal site
to 12$^{\circ}$ for a truncated icosahedron)
for the fullerene clusters. 
We compare the results of the present
work with  those of Itoh {\it et al.},$^{16}$ obtained with
the Harris functional}
\begin{tabular}{lcccccccc}
&\multicolumn{4}{c}{This work}&\multicolumn{4}{c}{Itoh {\it et al.}$^{16}$}\\
      & $\bar{r}$ (\AA)& $\sigma_s/\bar{r}$ & $\sigma_{m}/\bar{r}$ & $\phi$ &
$\bar{r}$ (\AA) & $\sigma_s/\bar{r}$ & $\sigma_{m}/\bar{r}$ & $\phi$ \\
\hline 
C$_{60}$& 3.59&0.000&0.000& 12.0$^{\circ}$ & 3.55&0.000&0.000& 12.0$^{\circ}$\\
C$_{240}$&7.18&0.023&0.027&8.5$^{\circ}$&7.06&0.021& 0.028 &7.9$^{\circ}$\\
C$_{540}$&10.69&0.038&0.054&9.6$^{\circ}$&10.53&0.033& 0.053 &9.2$^{\circ}$\\
\end{tabular}
\label{table:fullerenes}
\end{table} } }

As an example of a system with partially ionic character,
and with atoms with compact orbitals, we have performed
calculations on the $\beta$ phase of ${\rm C_3N_4}$, which
was proposed as a potentially very hard material by Liu and
Cohen.\cite{liu}
The calculations were done in supercells of 42 and 224 atoms,
with nearly identical results. A cutoff of 200 Ry for the 
charge density grid was used.
We obtain an accuracy 
better than 1\% in both the lattice constants and the
several inequivalent bond lengths, and 10\% in the bulk modulus, 
compared to other LDA calculations.\cite{liu}
These results contrast with those of the non-selfconsistent
Harris functional, which yield errors of 5\% and 16\%
for the distances and bulk modulus, respectively,
showing that selfconsistency is essential to obtain
reliable results in this partially polar system.

We have applied our method to study the 
structure of large, single shell icosahedral fullerene clusters. 
These are important to understand the observed sphericity of
multishell fullerenes (buckyonions).
For the single shell clusters, elasticity theory, as well as
empirical potential calculations, predict markedly
polyhedral shapes. Calculations performed by Itoh {\it et al.},\cite{itoh}
using the Harris-functional order-$N$ method,\cite{ordejon}
agree qualitatively with these results.
However, similar non-selfconsistent calculations\cite{yang} 
predict that even the large clusters are spherical. 
Here we have repeated the calculations with selfconsistent LDA
using the present method, thus improving over the non-selfconsistent 
nature of the former calculations.
Using a dynamical quenching algorithm, 
we have computed the equilibrium structure of three icosahedral
fullerene clusters: C$_{60}$, C$_{240}$ and C$_{540}$.
%The initial coordinates in the relaxation were taken
%from Itoh {\it et al.}.\cite{itoh}
A supercell geometry was assumed, with a cubic cell with sides of
12 \AA \hspace{0.06truecm} for C$_{60}$, 22 \AA \hspace{0.06truecm}
for C$_{240}$ and 34 \AA \hspace{0.06truecm} for C$_{540}$.
The calculations were done using a
cutoff of 100 Ry for the representation of the charge
density in reciprocal space,
and a different localization 
radius for $\sigma$ and $\pi$ type Wannier functions (2.5 and 4.0 \AA,
respectively).\cite{itoh} 
%For C$_{60}$ we obtain  bond lengths for the
%double and single bonds of 1.415 \AA \hspace{0.2truecm}
%and 1.472 \AA, respectivly,
%whereas the experimental values [9] for solid C$_{60}$ are 
%1.40 $\pm$ 0.015 and 1.45 $\pm$ 0.015 \AA, respectively.
%Other LDA calculations [10] yield values between 1.38 and 1.40  \AA
%
Increasing the localization radius to 4.1 \AA\ (both for $\sigma$ and $\pi$),
and/or increasing the grid cutoff to 150 Ry in the simulations 
changes the final relative distances in less than 0.4\%.
The results are summarized in Table 
\ref{table:fullerenes}. We see that our results are very similar
to those obtained by Itoh {\it et al.}, and confirm that, except
for C$_{60}$, the single shell clusters tend to be polyhedral,
instead of spherical, and that this polyhedral character
is more pronounced as the cluster size increases.

In conclusion, we have presented an efficient method
for selfconsistent LDA calculations with linear scaling.
We have analyzed the performance versus system size and
grid cutoff, and shown that simulations of systems
with hundreds of atoms are possible with small workstations.
This should open the possibility of very large scale
{\it ab initio} simulations in the near future.

%acknowledgments
We acknowledge R. M. Martin and Paul von Allmen for
many useful discussions, and D. A. Drabold and O. F. Sankey
for allowing us the use of many of their codes.
P. O. is indebted to
R. M. Martin and J. B. Adams for continuous support and 
encouragement.
This work was partially supported by 
DOE Grant No. DEFG 02-91ER45439 and 
DGICYT (Spain) Grant No. PB92-0169.


\begin{references}
\bibitem{minimi} The variables of the minimization are the
coefficients $c_{\mu i}$ of the expansion of the
localized wavefunctions in terms of the atomic orbitals: 
$\psi_i = \sum_\mu c_{\mu i} \phi_\mu$.
\end{references}
\end{document}